\documentclass[letterpaper]{article}
\usepackage{aaai24}

\usepackage{natbib}
\usepackage{times}
\usepackage{helvet}
\usepackage{courier}
\usepackage[hyphens]{url}  
\usepackage{lipsum} 
\usepackage{xspace}
\usepackage{amsfonts}
\usepackage{graphicx}
\usepackage[dvipsnames]{xcolor}
\usepackage{tikz}
\usepackage{pgfplots}
\usetikzlibrary{pgfplots.statistics}
\usepackage[ruled,vlined,linesnumbered]{algorithm2e}
\usepackage{subcaption}
\usepackage{verbatim}
\usepackage{amsmath}
\usepackage{longtable}
\usepackage{booktabs}

\usetikzlibrary{decorations.pathreplacing,calc}
\usetikzlibrary{decorations.pathreplacing,calc}
\usetikzlibrary{intersections}
\usepackage{soul}
\usepackage{pgf}
\usetikzlibrary{shapes.geometric}
\usetikzlibrary{shapes.arrows}
\usetikzlibrary{shapes,arrows,automata,positioning,fit,arrows.meta,calc}
\usetikzlibrary{decorations.markings}
\usetikzlibrary{patterns}
\usetikzlibrary{3d}
\usetikzlibrary{quotes,matrix}

\usepgfplotslibrary{fillbetween}
\pgfplotsset{compat=newest, 
	tick label style={font=\scriptsize},
	label style={font=\scriptsize},
	legend style={font=\scriptsize}}

\newcommand{\problem}{\ensuremath{\textsc{QBD-RankedDataGen}}\xspace}

\title{\problem: Generating Custom Ranked Datasets for Improving Query-By-Document Search Using LLM-Reranking with Reduced Human Effort 
}
\author{Sriram Gopalakrishnan, Sunandita Patra}
\affiliations{J.P. Morgan AI Research}        

\begin{document}

\maketitle

\begin{abstract}

The Query-By-Document (QBD) problem is an information retrieval problem where the query is a document, and the retrieved candidates are documents that match the query document, often in a domain or query specific manner. This can be crucial for tasks such as patent matching, legal or compliance case retrieval, and academic literature review. Existing retrieval methods, including keyword search and document embeddings, can be optimized with domain-specific datasets to improve QBD search performance. However, creating these domain-specific datasets is often costly and time-consuming. Our work introduces a process to generate custom QBD-search datasets and compares a set of methods to use in this problem, which we refer to as \problem. We provide a comparative analysis of our proposed methods in terms of cost, speed, and the human interface with the domain experts. 
The methods we compare leverage Large Language Models (LLMs) which can incorporate domain expert input to produce document scores and rankings, as well as explanations for human review. The process and methods for it that we present can significantly reduce human effort in dataset creation for custom domains while still obtaining sufficient expert knowledge for tuning retrieval models. We evaluate our methods on QBD datasets from the Text Retrieval Conference (TREC) and finetune the parameters of the BM25 model --which is used in many industrial-strength search engines like OpenSearch-- using the generated data.

\end{abstract}

\section{Introduction}

\begin{figure*}[ht]
    \centering
    \includegraphics[trim=0in 0in 0in 0in, clip,width=\textwidth]{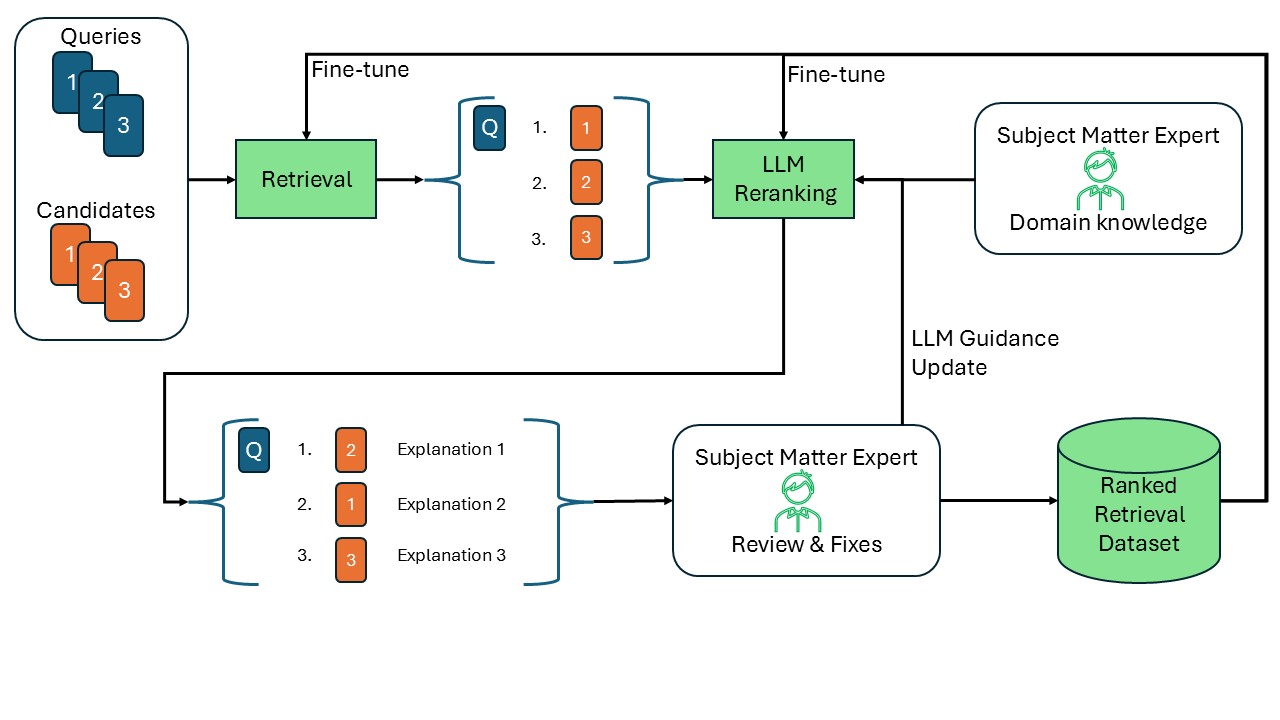}
    \caption{Flow of Information for QBD Dataset Generation with LLM and Subject Matter Expert}
    \label{fig:qbd_dataset_gen}
\end{figure*}


The Query-By-Document (QBD) problem is an Information Retrieval (IR) task that involves finding documents in a corpus that match a given query document. This task is particularly important in areas like law, where case matching can help decision making. It can also help with scientific literature review, or more generally any setting where document or case matching is important. Generating a QBD dataset for a target domain or distribution of documents \textit{with} expert human guidance or review can help with better selection or tuning of models for that target domain. However, the human cost and effort can make it challenging and expensive. This is especially so for very specific domains like a particular company's documents or legal cases; in such a scenario the expert human input cannot easily be obtained through crowd sourcing due to data-protection requirements, limited availability of domain experts, or for intellectual property protection. For such a dataset, the query-candidates mapping may include company-specific matching criteria or processes that an off-the shelf model for retrieval or re-ranking may not capture. A human-validated dataset allows for the selection or optimization of models for QBD search and candidate re-ranking. An additional motivation for our work is that there is a notable lack of such \textit{ranked}-datasets for QBD, particularly for long-form documents for QBD (where both query and document are long). 

Our work focuses on the challenge of generating ranked-datasets for QBD search which we call \problem. We describe and develop methods for the framework in Figure \ref{fig:qbd_dataset_gen} for use in \problem; the same process can apply for any ranked-dataset generation for retrieval tasks. The important steps in the dataset-generation process are the retrieval step, re-ranking step, and human-evaluation step (feedback). In this work, we focus on the reranking-step which is one of the steps that can reduce human effort in dataset generation, and in which expert knowledge can be effectively inserted (in turn reducing subsequent human effort in evaluation). Specifically, we compare different methods and models for the reranking step in \problem, especially those methods that can facilitate easier and better human review.

One of the benefits of having a curated dataset for a particular use-case is that the model for retrieval, as well as the model and method for re-ranking can be chosen and optimized with the dataset. In line with this thinking, we optimize BM25's parameters using the reranked results generated by the different methods we present in this work. This is to analyze how much better it can perform in the retrieval step on the datasets in question. We chose BM25 for its capability to handle both short and long documents easily and quickly in the retrieval step; many industrial-strength tools like Elasticsearch~\cite{Elasticsearch} utilize it. Working with long-documents is not an uncommon scenario with QBD search; long-documents can be seen when matching patent-filings or legal documents.

Our contributions are the following:
\begin{itemize}
    \item A process for generating \textit{ranked-datasets} for QBD that emphasizes the reduction of human expert effort with methods for the re-ranking step.
    
    \item A comparison of a set of models for the reranking step of the process on Text Retrieval Conference(TREC) datasets that fit the QBD problem setting. 
    
    \item Evaluation of BM25 after finetuning the BM25 parameters on the results of the reranking step for each of the different methods that we present.

    \item We discuss the tradeoff of different methods to reranking with respect to their human cost using information from human factors research.


\end{itemize}

\section{Related Work}
In this section, we review the existing QBD search datasets and discuss their limitations. We also describe retrieval techniques, LLM re-ranking and how domain knowledge can be used to finetune them.  
To our knowledge, there are no systems or papers that have explored dataset generation for the QBD problem. However, work has been done on the components we presented in Figure \ref{fig:qbd_dataset_gen}. Our literature review is summarized in Table~\ref{tab:relevant_areas}.

While many existing works \cite{wang2024doctabqa, choi2017coarse} focus on targeted answers or long retrieved documents with short, homogenous queries, our framework aims to provide an approach that allows for incorporating human domain experts in a sparse manner for generating high-quality datasets for QBD. Such datasets in turn can facilitate the fine-tuning of retrieval models and improving the re-ranking performance of the candidates returned. We focus our discussion on related work on the re-ranking step as that is the focus of this paper.

\begin{table*}[h]
\centering
\renewcommand{\arraystretch}{1.5}
\begin{tabular}{|p{5cm}|p{6cm}|p{4cm}|}
\hline
\textbf{Related area of work} & \textbf{Papers in the field} & \textbf{Our Contribution}\\ \hline
LLM Reranking  & \cite{zhu2023large}, \cite{pradeep2021expando}, \cite{liu2024demorank}, \cite{cho2023discrete}, \cite{qin2023large}, \cite{luo2024prp}, \cite{pradeep2023rankvicuna}, \cite{baldelli2024twolar}, \cite{rashid2024ecorank} & We use LLM Reranking for QBD dataset generation.\\ \hline
QBD Datasets & \cite{risch2020patentmatch}, \cite{macavaney2021simplified}, \cite{zhou2024sst}, \cite{he2024match}, \cite{askari2023closer}, \cite{askari2024retrieval} & We reduce the human effort needed to construct QBD datasets.\\ \hline
Ranked Dataset Generation  & \cite{awan2021top}, \cite{zhang2018semre}, \cite{wang2016identifying}, \cite{frej2019wikir}, \cite{danilevsky2014automatic} &
We generate a ranked dataset for long queries and documents.\\ \hline
Retrieval Methods for Long Documents & \cite{saad2024benchmarking}, \cite{abolghasemi2022improving}, \cite{cohan2020specter}, \cite{askari2023closer}, \cite{askari2024retrieval}, \cite{mansour2024revisiting}, \cite{varadarajan2006system}, \cite{zhuang2023rankt5} &
We provide a way to construct a diverse set of evaluation benchmarks for QBD with reduced human effort.
\\ \hline
LLM Based Finetuning of Retrieval Methods & \cite{jeronymo2023inpars}, \cite{bonifacio2022inpars}, \cite{dai2022promptagator}, \cite{thakur2021heterogenous}, \cite{nogueira2020document} &
We provide a way to finetune retrieval QBD methods using LLM re-ranking.\\ \hline
Human Feedback Based Finetuning for Specialized Use Cases & \cite{nouriinanloo2024re}, \cite{askari2023closer}, \cite{wang2025bring}, \cite{long2024llms}, \cite{guan2023cohortgpt}, \cite{xu2023knowledge}, \cite{anisuzzaman2025fine} &
We provide a way to use human expert knowledge for QBD dataset generation.\\ \hline
\end{tabular}
\caption{Summary of literature review showing the related areas of work, the corresponding papers and our contribution with respect to each area.}
\label{tab:relevant_areas}
\end{table*}

\subsection{LLM Reranking}

LLM re-ranking has been used for successfully for multiple applications with short documents/phrases. A recent summary of the major re-ranking strategies \cite{zhu2023large} shows that existing work does not focus on the QBD search problem, much less dataset generation for QBD with or without human experts. There are problem-specific challenges to using LLMs for re-ranking in QBD dataset generation and when involving humans in the loop that merit a focused analysis.  In this work, we chose to adapt two of the re-ranking strategies reviewed by \cite{zhu2023large}; the pointwise method and the pairwise method. By leveraging LLM re-ranking techniques and the easy natural language interface that LLMs provide, we aim to evaluate and adapt them for the QBD problem in a way that allows domain experts to be involved without taxing them. This can help generate \textit{ranked} retrieval datasets that are currently lacking for QBD search, and especially for long-document QBD search.

\subsubsection*{Supervised LLM re-ranking}
LLM's pretraining phase usually lacks ranking awareness. Supervised finetuning can improve pre-trained large language models (LLMs) for re-ranking. Supervised re-rankers can be broadly categorized into three types based on their architecture: encoder-only, encoder-decoder, and decoder-only. 

Encoder-only models transform query and document(s) to relevance scores. For example, the Duo component in \cite{pradeep2021expando} takes in a query with a pair of candidate documents and returns a token indicating which candidate is more relevant ("true" if document 1 is more relevant). This is analogous to what we do in pairwise comparison, albeit with a powerful LLM prompted to return the token as part of the response. 

Encoder-decoder models \cite{zhuang2023rankt5, pradeep2021expando} treat document ranking as a generation task, optimizing models to generate classification tokens for relevance. Decoder-only models \cite{zhang2024two, ma2024fine}, format query-document pairs into the prompts and use two-stage training to enhance ranking performance.

\subsubsection*{Unsupervised LLM Reranking}

Finetuning can result in improved re-ranking performance. However, this requires a domain specific dataset with which to fine-tune and we have found very few datatsets for QBD search, and may not be there for all applications; this is the problem that this paper targets. 

Another problem with finetuning an LLM is that effectively finetuning LLMs can be difficult in practice. To overcome this, recent works have successfully used unsupervised prompting strategies for re-ranking tasks. These can be broadly categorized into pointwise, pairwise, and listwise methods. In this work, we do unsupervised re-ranking of documents using the pointwise and pairwise methods. We avoid the listwise method as we observed re-ranking performance can vary appreciably with order of documents in the prompt. Another problem is that if the documents are long, then we may not be able to fit all the documents into the prompt due to context length limitations of the LLM.

In pointwise methods, the LLMs judge the query-document relevance individually. One approach is to evaluate and generate relevance scores such as in DemoRank \cite{liu2024demorank} using demonstrative examples. Another is custom query generation, such as Co-Prompt \cite{cho2023discrete} which works by improving the prompt. 

On the other hand, pairwise methods involve comparing document pairs to determine relevance, using aggregation methods like AllPairs \cite{qin2023large} (which is similar to what we do in the pairwise comparison method) and approaches like graph-based aggregation \cite{luo2024prp}, though they face high time complexity. None of these works consider the human-costs during in the process or consider the problem of dataset generation for subsequent optimization of retrieval or re-ranking models.


\cite{pradeep2023rankvicuna} introduced RankVicuna, a zero shot re-ranking method evaluated on TREC datasets \cite{craswell2020overview}, which consist of small documents. They develop a 7B parameter model by instruction finetuning of open source LLMs. This uses a sliding window approach to handle a large number of documents to re-rank.  Using such a model would require training data to finetune it, which is the problem we are trying to address with this work.


\cite{baldelli2024twolar} proposed TWOLAR, a two-step LLM-augmented method for passage re-ranking. This method introduces a scoring strategy and a distillation process to create a diverse training dataset outperforming state-of-the-art models with significantly fewer parameters. They highlight the reliance of existing methods on costly human-annotated labels which they bypass by using an LLM. However, LLMs may not fully bypass the need for human expert guidance as specific re-ranking tasks can be challenging for them as we show in our experiments.

\subsection{QBD Datasets}

\cite{macavaney2021simplified} provides a synthetic dataset of expert generated patient descriptions to match against documents which are clinical trial descriptions. They evaluate the ranked list generated by the participating team by presenting top few results to physicians trained in medical informatics. We use this clinical trials dataset in our experiments as it fits the problem of QBD search where the query is a document describing a potential (synthetic) patient's medical record and the candidates are documents describing clinical trials.  

The need for high-quality datasets in QBD is further emphasized by \cite{zhou2024sst}, who developed a score-based solution for the QBD problem, evaluated on two Chinese datasets. They note the lack of publicly available datasets for long document matching tasks, underscoring the importance of our work to help address this problem. Similarly, \cite{he2024match} evaluated their QBD solution on Chinese news datasets, reinforcing the necessity of diverse and comprehensive datasets for QBD \cite{askari2023closer, askari2023generating, askari2024retrieval}.
To our knowledge, there is no work on  QBD dataset generation that focuses on  reducing the human effort in generating QBD datasets.

\subsection{Ranked Dataset Generation}
Ranked Dataset generation has been explored for short documents and keyphrases in the data mining and retrival community \cite{awan2021top, zhang2018semre, wang2016identifying, frej2019wikir}. \cite{danilevsky2014automatic} introduced a framework for topical keyphrase ranking for a collection of short documents. Their ranking function compares the quality of keyphrases, and is evaluated by humans. No work to our knowledge explores how humans and LLMs can be effectively combined for generating ranked datasets for QBD search.

\subsection{Retrieval methods for long documents}

Retrieval methods are well developed for shorter documents and for use-cases like question-answering. However, this step is still challenging for two scenarios; when the documents are very long; and when retrieval requires query-specific or task-specific search, and the retrieval model has not been trained for that type of search. The latter  can happen when there is a paucity of data for that use case.

Retrieval for long documents is challenging because it's difficult to embed long contexts during both the pretraining and finetuning of the base model, especially with batch size limitations. Additionally, evaluating these models is tough due to a lack of available benchmarks. In \cite{saad2024benchmarking}, the authors introduce an 80 million parameter model called M2-BERT, which is designed for retrieval tasks. This model can efficiently handle longer input contexts by using Monarch matrices, which allow for faster processing of both the input sequence length and the model's dimensions. This approach helps in managing larger inputs without a significant increase in computational complexity.

\cite{abolghasemi2022improving} processes a technique to improve BERT's retrieval effectiveness by refining it's document representations to incorporate the context for long queries. They evaluate on the SciDocs dataset of academic papers \cite{cohan2020specter}. The line of work, \cite{askari2023closer}, \cite{askari2023generating} and \cite{askari2024retrieval} for long legal case retrieval and ranking present efficient longformer based architecture that achieve state of the art performance on the legal dataset from \cite{askari2023closer}. Another approach to solving the QBD problem involves a two-step process: first, summarizing the query as suggested by\cite{mansour2024revisiting, varadarajan2006system}, and then performing retrieval using the summarized query. 

In \cite{zhuang2023rankt5}, RankT5 is introduced for text ranking that directly output ranking scores and can be fine-tuned with pairwise or listwise ranking losses. The experiments demonstrate that their models achieve significant performance improvements on text ranking datasets, with listwise fine-tuning providing superior zero-shot ranking performance on out-of-domain data compared to classification-based fine-tuning.

Retrieval approaches like these can benefit from a diverse range of datasets, which can be generated to improve model training and evaluation. This is one of the main goals of our work.

\subsection{LLM Based finetuning of retrieval methods}

Fine-tuning retrieval methods using the output of large language models (LLMs) is an effective approach to improve the performance of retrieval. This has been done in practice for small query and document sizes. The InPars and Promptagator methods, as proposed in  \cite{jeronymo2023inpars, bonifacio2022inpars}, leverages LLMs to generate synthetic query-document pairs through few-shot examples, which are then utilized to train retrieval models. \cite{bonifacio2022inpars, dai2022promptagator} demonstrated the potential of LLM-generated synthetic data in retrieval tasks, and introduces a re-ranker as a filtering mechanism to refine the selection of synthetic examples, aiming to improve retrieval effectiveness.  Building on this, InPars-v2 \cite{jeronymo2023inpars} uses open-source LLMs and re-rankers to select relevant synthetic pairs for training, achieving competitive results on the BEIR benchmark \cite{thakur2021heterogenous}. The method incorporates a BM25 retrieval pipeline followed by a monoT5 re-ranker \cite{nogueira2020document} fine-tuned on InPars-v2 data. Synthetic data can be helpful for QBD search as well if the document generation process can be sufficiently described to cover the distribution of documents. Our method comes in after the corpus of documents has been decided.

\subsection{Human feedback based finetuning for specialized use cases}

Use of human feedback to improve re-ranking methods is limited. 
\cite{nouriinanloo2024re} uses a pre-filtering step with a small number of human-generated relevance scores combined with LLM relevance scoring to effectively filter out irrelevant passages, thereby enhancing the re-ranking performance of smaller models.
\cite{askari2023closer} does expertise aware post training in their longformer architecture introduced for legal case retrieval. However, domain expert knowledge has been widely used in finetuning LLMs successfully in other NLP tasks, such as, text classification and question answering by LLMs. \cite{wang2025bring} and \cite{long2024llms} provide surveys of such methods. 

\cite{guan2023cohortgpt} finetunes an LLM using a human expert constructed knowledge graph and chain of thought prompting. They observe better performance in disease classification from systems. and is intended to help with selecting candidates for randomized control trials. 
\cite{xu2023knowledge} uses context-informed human prompting to improve clinical knowledge extraction by LLMs, improving performance and diversity.  
\cite{anisuzzaman2025fine} focuses on finetuning LLMs with human expert knowledge for three use cases in the medical field, medical research assistance, clinical decision support, and patient interaction automation. One of the limitations of these approaches is the cost of gathering human expert knowledge to do the finetuning. Our approach seeks to reduce this tax on the human expert when obtaining data to tune the retriever and re-ranker.



\section{Problem Formulation}

\subsection{Input}
\begin{itemize}
    
    \item \textbf{($Q$, $D$)}: The dataset of query  documents and candidate documents respectively. The documents maybe real documents or synthetically generated documents that conform to domain knowledge.
    
    \item \textbf{$\mathbf{t}$}: Desired number of query-candidate ranked sets to be generated for training or test.
    
    \item \textbf{Ranked-List Filter Function} \\ \( f_r: \{(q, d_1, d_2, \ldots, d_n)\} \to \{0,1\} \)\\
    This function takes a set consisting of one query document $q \in Q$ and zero or more candidate documents $(d_1, d_2, \ldots, d_n)$ as input, with $d_i \in D$. It outputs a binary value indicating whether the query document and its associated candidate list form a valid data point.

    \item \textbf{Oracle}: This is the subject matter expert who can provide or update instructions or rules for re-ranking candidates, can review and correct the rank order after the model has re-ranked, and optionally reject a data point as invalid. 

\end{itemize}

The query and candidate datasets are separate as the documents maybe heterogenous (different types of documents). An example of this will be seen in the clinical trials dataset used in experiments.

The ranked-list filter function is configured by the user, and can be used to remove out those queries that have too few relevant matches in the dataset after the retrieval step. This is for generating a helpful ranked list of candidates; helpful here means has sufficient information or signal to be used to tune or select a model for subsequent QBD search. For example, if there is only 1 candidate match and the user would like to finetune a model using a contrastive-loss function, then atleast two candidates would be needed per query. This filter can also be used to avoid having too many candidates for a query as the cost during the re-ranking step maybe prohibitive; LLM use can be expensive, and human cost of validating many pairs of text can be a problem too.

The oracle in our problem formulation is typically a human, whose time or usage we would like to reduce. The human's effort is expensive, both in terms of cost and speed. The oracle could also be an expensive LLM model controlling/instructing a cheaper and faster re-ranking LLM model.  


\subsection{\problem  Objective}

The objective is to map queries selected to a ranked list of candidate documents. This is defined as:

\begin{align*}
R: Q \times Z^+ &\to \mathcal{P}(D) \\
R(q, t) &= \langle d_1, d_2, \ldots, d_m \rangle \\
\text{where } \{d_1, d_2, \ldots, d_m\} &\subseteq D, \, m \leq t
\end{align*}

where $R$ is the retrieval and reranking process, and $Z^+$ is the set of all positive integers.. Additionally, each query to candidates-list should pass the ranked-list filter function. The generated query to ranked-list mapping can then be used for selecting models for QBD search, or optimizing models. We will refer to this as the generated training signal for simplicity (even if it can also be used for selecting models rather than training or optimizing them).

\section{Methodology}

In this section we will describe the process in Figure \ref{fig:qbd_dataset_gen} and in particular focus on the re-ranking process.

For obtaining the training signal for a QBD document set, we build on a typical retrieval and re-ranking pipeline that is popular in IR. In the retrieval stage, a typically fast but less accurate method is used to collect a set of possible matching candidates for a given query document. This is followed by the re-ranking stage where a more computationally intensive process re-ranks the candidate documents to better fit the query. Our work focuses on the re-ranking stage since this is where human-knowledge and feedback can be more easily incorporated. Since our main focus is on the re-ranking step with LLMs, we do not compare different retrieval methods in this work. Retrieval methods include approaches like BM25, vector embeddings, or hybrid approaches. The retrieval method can certainly affect the quality of datasets generated; however, to better evaluate the reranking methods (which is the focus), we remove the confounding effect of retrieval methods and assume the re-ranker gets the correct top-k documents, and only has to rank them to the correct order.

In this work, we compare different approaches to using LLMs for re-ranking, as some modes of presenting information are easier for people to work with than others (discussed in subsequent section). The cost of using different LLM re-ranking approaches can be seen in re-ranking performance and computation costs. We sought to investigate different tradeoffs with LLM re-ranking methods that can involve a human in the loop.

The methods we present are distinguished by,

\begin{itemize}
    \item method of re-ranking.
    \item how domain expert knowledge is inserted before the start of re-ranking.
    \item how the method presents information and (optionally) explanations to a domain expert after the re-ranking process; the explanations can help the expert validate or correct the decisions made by the LLM for the generated dataset.
\end{itemize}

The re-ranking methods are categorized into single-document scoring methods (pointwise methods), and pairwise-document scoring methods. We do not do listwise-document scoring (more than 2 documents) for two reasons: The first is that empirical data showed that performance drops with more than 2 documents, and even with two we noticed the order of documents could affect the result; the second is that if one is working with long documents, then more than two can exceed the maximum context length for many LLMs.

\subsection{Single-Document Scoring Methods}

As the name implies, these methods consider each candidate document (with a given query) separately and score it. We prompt the LLM to output a score between $[0,1]$ as a measure of how good of a fit the candidate is to the query document.  Any expert guidance can be inserted into the prompt on how to evaluate the candidate. We experiment with and without expert knowledge inserted in the prompt as well.

All scoring methods that use an LLM are also asked to output an explanation for their decisions. This has a dual benefit of simulating reasoning in the LLM inference, as well as helping update the expert to update the knowledge inserted in the prompt or correct the decision in the dataset. The models and variations on this method are detailed in the experimental section.




\subsection{Pairwise-document Scoring Methods}

These methods are analogous to the Single-document scoring method, albeit the scores are assigned to comparisons of pairs of documents. A score of +1 is given to the document that matches better and -1 if it is the worse of the pair. 0 is given if the LLM determines that neither document is necessarily a better match give the information it is given.

LLMs can be sensitive to the order in which information is presented. This was also noticed in the work \cite{qin2023large}. So we present the documents in both orders as the authors of the aforementioned work did. We also compare all pairs of documents to improve the overall ranking as was also done in \cite{qin2023large}.
The different pairwise methods that we experimented with is in the experimental section under the re-ranking step.


\subsection{Measures}

To evaluate the ranked candidates at the end, we use the following measures:
\\[1em]



\textbf{Kendall's $\tau_{b}$ Coefficient}: This is a variant of Kendall's Tau Coefficient that can handle ties when comparing two ranked-lists. Given two ranked lists side-by-side, we get tuples of ranks at the same position; for example the first position could have (1,3) and the second position could have (2,1). If a pair of tuples agree in the relative  order of ranks then it is concordant, such as (2,4) and (3,7). If they disagree it is discordant; for example (1,3) and (2,1) are discordant.
For a pair of tuples, if the rank is the same in the first variable, then it is a tie in the first variable; similarly for the second variable.
Given this information, we can compute the $\tau_{b}$  coefficient as 

\begin{equation}
    \tau_b = \frac{C - D}{\sqrt{(C + D + T_1)(C + D + T_2)}}
\end{equation}

where C is the number of concordant pairs of all possible pairs of tuples; D is the number of discordant pairs; $T_1$ is the number of pairs with ties in the first variable(from the first list) only, and $T_2$ for the second variable only. 
\\[1em]
\textbf{Precision@K}: The ratio of relevant documents in the top-k candidates retrieved, where relevance is determined based on whether the candidate is in the ground truth.
\\[1em]
\textbf{Mean Average Precision (MAP)}: For each query we compute the average precision, which is average of the Precision@K for all possible positions in the retrieved list. Then we take the mean over all queries to compute MAP,
\begin{equation}
    \text{MAP} = \frac{1}{|Q|} \sum_{q=1}^{|Q|} \left( \frac{1}{R_q} \sum_{k=1}^{N_q} \text{Precision}(k) \cdot \text{rel}(k) \right),
\end{equation}
where $|Q|$ is the total number of queries; $R_q$ is the number of relevant documents for query $q$. $N_q$ is the number of documents returned for query ( q ).
(\text{Precision}(k)) is the precision@K; and (\text{rel}(k)) is an indicator function that is 1 if the item at position ( k ) is relevant, and 0 otherwise; \text{rel}(k) ensures that irrelevant documents returned are not counted towards the average precision computation in the inner for loop.
\\[1em]
\textbf{Spearman's Rank-order Correlation}: is a measure of rank-correlation by giving the direction and magnitude of the monotonic relationship between two variables. 
\begin{equation}
     \rho = \frac{\sum (R_X - \overline{R_X})(R_Y - \overline{R_Y})}{\sqrt{\sum (R_X - \overline{R_X})^2} \sqrt{\sum (R_Y - \overline{R_Y})^2}} 
\end{equation}
\\[1em]
\textbf{Mean Reciprocal Rank}: is mean of the reciprocal of the rank at which the first relevant document appears. This is defined as follows:
\begin{equation}
     \text{MRR} = \frac{1}{|Q|} \sum_{i=1}^{|Q|} \frac{1}{\textit{rank}_i} 
\end{equation}
where $|Q|$ is the number of queries and $\textit{rank}_i$ is the rank at which the first relevant document appears.
\\

\section{Experiments}


\subsection{Datasets}

For our experiments we used the TREC clinicaltrials 2021 datset~\cite{trecclinicaltrials}, and the TREC CORD-19 dataset~\cite{voorhees2021treccovid}. The datasets were downloaded using the IR datasets library~\cite{IRdatasets}.
For experiments on QBD-search, we wanted queries or candidates to be longer documents than the short snippets that we find in Question-Answer datasets. We could not use datasets where the answer is in a long document, but only a small section of the entire document is relevant for the query such as the MS-MARCO dataset~\cite{msmarco}, as the document is not the match, but a snippet in the document; this is not the QBD search problem which is matching documents. Additionally, for our experiments we needed a ranked list of candidates, and not just a match/not-match label since we are focusing on the reranking step in the process described in Figure \ref{fig:qbd_dataset_gen}.

We chose IR datasets collection as it is one of the few we found with datasets that have the ground-truth containing matching scores beyond just 0 or 1 (match or not) for each query-candidate mapping, and enough queries with more than 3 matching candidates;  having more than 3 helps to get a meaningful ranked list of candidates per query. Additionally some of the documents in the provided query-candidate mappings are long documents in these datasets which can happen in QBD search. We could not find other satisfactory document-retrieval datasets with human validated ranked lists. The lack of datasets is one of the problems that we are trying to help with in this work. 

\subsubsection{TREC Clinicaltrials 2021:}
The Text Retrieval Conference (TREC) clinicaltrials dataset~\cite{trecclinicaltrials} is a dataset for the task of matching synthetic patient medical descriptions (written by experts) to real clinical trial descriptions. The trial descriptions have inclusion and exclusion criteria that make the matching task harder without non-trivial semantic understanding. The dataset evaluates retrieval methods by categorizing outcomes into eligible, excludes, and not relevant which was scored as 2, 1, and 0. There were only 75 queries with scored candidate lists that could then be ranked, but each query had on average 75 candidates with score 2, 80 with score 1, and 323 with score 0. The total number of query-candidate pairs is 35,832. 


\subsubsection{CORD-19 TREC COVID:}
In this dataset the queries are descriptions of topics related to covid 19 and the candidates are full covid-19 research papers. The candidates are scored as 2, 1 or 0 based on the degree of relevance. There are 50 queries in total, but like with the clinical trials dataset, each query has a large number of candidates in it's scored list. Each query has on average 312 candidates with score 2, 221 candidates with score 1 and 853 candidates with score 0.\\

Each of the aforementioned datasets is split into training and test set. Training-set is a bit of a misnomer here as we are not really training a model with it, but using it to generated a dataset. We say training set, because the signal generated from it (in the dataset made) is used to optimize the parameters of a bm25 model. The optimized bm25 model is then tested on a separate test set. 

From each query we take up to 10 candidate documents per score; for example 10 candidates with score 2, 10 candidates with score 1 and 10 candidates with score 0. We first remove 20 percent of the queries as pure test set queries. For the remaining queries, we convert them to query and single-candidate pairs and randomly shuffle them. We randomly select 100 pairs, from which we drop queries that have only one candidate from this selected list (goes to the test set). The pairs left after this filtering is the training set, and all other pairs are the test set. We intentionally have a small training set to reflect the lower number of data points a domain expert in a company maybe willing to give. 

Using this process on the clinicaltrials dataset resulted in 65 query-candidate pairs; each query has 2 to 6 candidates. The test set is comparitively much larger for clinicaltrials dataset with 75 queries and 2144 candidates over all the queries' ranked list. No candidate documents are shared between the train and test sets. There are 15 queries that are only in the test set. The smaller \textit{training} set reflects the intended setting where a subject-matter(domain) expert is very expensive or can offer a limited amount of time for data annotation.

Similarly, for the Cord-19 dataset, the training set has 83 query-candidate pairs. The test set has 50 queries with 1419 candidates. There are 10 queries that are only in the test set. Note that in both the datasets, all of the training queries appear in the test set as well, but with different candidates in the ranked list for those queries. The other queries in the test set are unique to the test set. None of the candidate documents overlap between the training and test sets.


\subsection{Removing Retrieval Impact on Reranking}
Since the focus in this work was on the reranking step, we wanted to avoid any impact the choice of the retrieval method would have on reranking. So the documents that go into the reranked is a subset of the relevant documents as mentioned in the dataset, as well as some irrelevant (to the query) but related documents (as determined by the human evaluators for the TREC dataset). 


\subsection{Reranking Step}
For evaluating the reranking step, we input each query and a subset of candidate texts (without score or rank information) from the ground truth of each dataset to each reranking method; the task we care about is the reranking performance and how well it matches the ranking of the ground truth. There are some irrelevant documents that were from the ground truth (score of 0), but we did not include a lot of additional irrelevant documents as that could significantly increase the monetary costs of inference with some LLMs, and also the human cost to review in the full implementation of the framework in Figure \ref{fig:qbd_dataset_gen}.

For reranking, the baseline approach was ranking based on the score returned by bm25 with the query text with default parameters k1 = 1.5 and b = 0.75. The bm25 indexing was only on the training set of documents to make it a more fair comparison with the other methods; this is as opposed to indexing a larger set of documents that are not relevant to the queries. This baseline was compared to Open-AI's text-3 large embedding model, and gpt-4o-mini for the reranking task. Each of these models was used in a set of methods for reranking. The set of methods are as follows:

\begin{itemize}
    \item Single Candidate Scoring with Embedding (\textbf{SCS-emb}): In this method we score each candidate for a query based on the embedding similarity. The scores are then used to rank the candidates 

    \item Single Candidate Scoring with LLMs (\textbf{SCS-llm}): We prompt the LLM to return a score between $[-1,1]$ based on how well the candidate matches the query (both are also in the prompt). After llm scoring, we rank based on the scores.
    
    \item Single Candidate Scoring with Instructions(\textbf{SCS-instr}): This is builds on SCS-llm; the prompt will also have instructions on how to match. The instructions would come from the oracle (which could be a human or advanced LLM). In our experiments the instructions are the task description accompanying the dataset given by humans.

    \item Pairwise Candidate Scoring with LLMs (\textbf{PCS-llm}): With this method, we prompt the LLM to return a score that is one of $\{-1,0,1\}$. These scores (respectively) correspond to the following scenarios: candidate 1 is a worse match than candidate 2; both candidates are equally good matches, equally bad, or the LLM is uncertain; candidate 1 is a better match than candidate 2. The prompt will contain the query and a pair of candidates. For each candidate text, we sum up all the scores from each pairwise comparison that it was involved in, and compute the total score which is then used to rerank.   
    
    \item Pairwise Candidate Scoring with Instructions (\textbf{PCS-instr}): This is builds on PCS-llm; the prompt will also have instructions on how to match. The instructions come from the task description accompanying the dataset.

\end{itemize}

If the scores for two candidates are equal, we used the competition ranking method of assigning ranks to handle conflicts. This means that if a subset of candidates have the same score, they receive the same rank, and the next candidate receives a rank that skips the number of tied candidates. For example, if two candidates are tied for first place, they both receive a rank of 1, and the next candidate receives a rank of 3, skipping rank 2.

We do not adopt reranking by putting all candidates into the LLM and asking for a ranking order for three main reasons. 
(1) In QBD, each document maybe long and inserting all candidates into the prompt can exceed context length of many LLMs, and LLMs that can handle very long contexts can have prohibitive costs.

(2) Another reason is that in our experiments, we noticed that changing the order of candidate documents in pairwise candidate scoring methods sometimes changed which document the LLM thought was a better match. If the reasoning could be affected by order, then adding all candidates could make it worse due the number of possible orders and variations in scoring. There is research that shows transformer attention maybe biased towards information near the start and end of the context, especially in long contexts~\cite{attention_bias_1,attention_bias_2}. Any such attention bias can have an impact on the ranking output, so we limit ourselves to at most comparing 2 candidates at a time.

(3) The last reason is because of human factors. It is easier for the human to review single candidate decisions and explanations, or pairwise comparisons; pairwise comparisons are even better as shown in human factors research and methods have been developed to handle pairwise comparisons in large datasets as well~\cite{pairwise_humans_bradley_terry}. On the other hand, asking a human to evaluate or correct a larger ranking of items would require the human to keep track of all possible ranking or scoring decisions at once. This can be a harder task based on the amount of information (all candidates' information) that the person must hold in memory while evaluating. 

Another work titled \textit{Pairwise Preference Search }(PairS)~\cite{pairwise_pref_with_llm} has also shown that LLMs themselves work better with pairwise comparisons to rank a list of candidates to align better with human judgments. This was done in the context of evaluating the quality of generated long-form content for human preferences, not for ranking long-form documents in QBD retrieval. It does however, give us reason to think that LLM performance in ranking long-form text can be done better with pairwise comparisons. 

In our experiments, after reranking with different methods, we use the signal generated (query-candidate ranking) to select/tune bm25 ${k1, b}$ parameters on the training set. We chose bm25 as it can easily handle long documents well, and is used in industrial-strength retrieval engines ~\cite{Elasticsearch}. The tuned bm25 was evaluated on the withheld test set for each dataset. This was compared against fine-tuning using the ground truth for those same query-candidate lists. The ground truth represents the result after an oracle (human expert) corrects the reranking from the reranking-model output. The comparison is to give us a sense of how much of a performance boost can we expect without human validation of the ranked results. The fine-tuning was done with the Optuna library~\cite{optuna} using 50 trials per experiment, and searching in the range [1.2,2.0] for $k1$ and [0.1,1.0] for $b$. The measure used for tuning was mean average precision.

\section{Results}

We divide the results into two sections: the evaluation of the re-ranking performance by different methods; and the evaluation of the BM25 model after parameter fine-tuning using the training signal from the previous step.

\subsection{Evaluating Reranking Performance}

We present the reranking performance in the tables \ref{tab:clinicaltrials_2021_trec_ct_2021_train} and \ref{tab:cord19_fulltext_trec_covid_train}, each method is further qualified in parenthesis by what underlying model is used in that method (eg: gpt4omini). 

For these two datasets, there is no overall best method in terms of ranking performance over all measures and both datasets. We expected the methods with instructions from the retrieval task description in the dataset to do better, but that was not seen in the data. Perhaps with better human instructions or adding well-thought-out domain-specific examples to the prompt, we could get better performance. This was not available for the datasets we used.

For the clinical trials dataset, one interesting note is that adding the instructions actually hurt the performance more than helped for both the pairwise and single score methods. The instructions came from the TREC description of the task, and it may not have been sufficient or confusing for the LLMs. There is no clear benefit seen for the cord19 dataset either. It is unlikely that adding instructions for matching will always hurt performance or not make a difference. We can only say that using the task description from TREC datasets in the instructions to the LLM was not helpful for reranking with the methods we used. Improving the prompt instructions is one of the reasons to have a human feedback path; this is in Figure \ref{fig:qbd_dataset_gen} between reviewing the reranking results to new reranking instructions.

A surprising result in our experiments was how well reranking by embedding similarity using embeddings from OpenAI's \textit{text embedding-3 large} model worked in the clinical trials dataset. This is likely because matching text and phrases may have been sufficiently informative for the clinical trials task. This is supported by the fact that BM25 without tuning also did relatively well for that dataset (Table \ref{tab:clinicaltrials_2021_trec_ct_2021_train}); the correlation measures in the table for BM25 are more than half of the llm based methods, whereas in the other dataset (Table \ref{tab:cord19_fulltext_trec_covid_train}) the ratio of performance measures is much less than half.

\subsection{Evaluating BM25 Fine-Tuned Performance}
In Tables \ref{tab:clinicaltrials_2021_trec_ct_2021_test} and \ref{tab:cord19_fulltext_trec_covid_test} we have the results from tuning the BM25 model's $\{k1,b\}$ parameters using data from re-ranking. The data used and associated names for the tuned BM25 are as follows:
\begin{itemize}
    \item BM25IdealTest: Tuned on the ground truth of the test data. This is to give a bound on the best that can be achieved.
    
    \item BM25IdealTrain: Tuned on the documents and ground truth of the training set. This would correspond to the Oracle(human) correcting the re-ranked results from the LLM or other models.
    
    \item BM25 SCS-llm(gpt4omini) : Tuned on the training set of documents with the ranking produced by the SCS-llm method using Gpt-4O-mini model and no instructions.

    \item BM25 SCS-instr(gpt4omini) : Tuned on the training set of documents with the ranking produced by the SCS-instr method using Gpt-4O-mini model with instructions.

    \item BM25 PCS-llm(gpt4omini) : Tuned on the training set of documents with the ranking produced by the PCS-llm method using Gpt-4O-mini model and no instructions.

    \item BM25 PCS-instr(gpt4omini) : Tuned on the training set of documents with the ranking produced by the PCS-instr method using Gpt-4O-mini model with instructions.

    \item BM25: this is the baseline model with parameters $\{k1 = 1.5 ,b = 0.75\}$ which is a standard default set of parameters for BM25.

\end{itemize}

In the results we see that only when BM25 is tuned with the ground truth (either training or test set), does the performance exceed the default parameter settings of $\{k1 = 1.5 ,b = 0.75\}$. This would imply that getting human validation and correction in the process of Figure \ref{fig:qbd_dataset_gen} maybe essential. Alternatively, it can be that for some datasets, the re-ranked data generated by LLMs could be sufficient for fine-tuning the retrieval model. However, this was not supported by our experiments. 

\onecolumn


\begin{longtable}{lrrrrr}
\caption{Results for clinicaltrials\_2021\_trec\_ct\_2021. Higher values indicate better performance.} \\
\toprule
Ranker Name & KendallTau & Spearman & MRR & Prec@k=3 & MAP \\
\midrule
\endfirsthead
\caption[]{Results for clinicaltrials\_2021\_trec\_ct\_2021} \\
\toprule
Ranker Name & KendallTau & Spearman & MRR & Prec@k=3 & MAP \\
\midrule
\endhead
\midrule
\multicolumn{6}{r}{Continued on next page} \\
\midrule
\endfoot
\bottomrule
\endlastfoot
PCS-instr(gpt4omini) & 0.541000 & 0.582000 & 1.000000 & 0.987000 & 0.997000 \\
PCS-llm(gpt4omini) & 0.683000 & 0.721000 & 1.000000 & 1.000000 & 1.000000 \\
SCS-emb(textembedding3large) & 0.703000 & 0.737000 & 0.980000 & 0.987000 & 0.986000 \\
SCS-instr(gpt4omini) & 0.590000 & 0.622000 & 0.980000 & 0.980000 & 0.983000 \\
SCS-llm(gpt4omini) & 0.676000 & 0.704000 & 1.000000 & 0.980000 & 0.993000 \\
bm25 & 0.397000 & 0.426000 & 0.960000 & 0.887000 & 0.948000 \\
\label{tab:clinicaltrials_2021_trec_ct_2021_train}
\end{longtable}

\begin{longtable}{lrrrrr}
\caption{Results for cord19\_fulltext\_trec\_covid} \\
\toprule
Ranker Name & KendallTau & Spearman & MRR & Prec@k=3 & MAP \\
\midrule
\endfirsthead
\caption[]{Results for cord19\_fulltext\_trec\_covid} \\
\toprule
Ranker Name & KendallTau & Spearman & MRR & Prec@k=3 & MAP \\
\midrule
\endhead
\midrule
\multicolumn{6}{r}{Continued on next page} \\
\midrule
\endfoot
\bottomrule
\endlastfoot
PCS-instr(gpt4omini) & 0.529000 & 0.568000 & 0.960000 & 0.900000 & 0.953000 \\
PCS-llm(gpt4omini) & 0.551000 & 0.583000 & 0.980000 & 0.920000 & 0.967000 \\
SCS-emb(textembedding3large) & 0.305000 & 0.350000 & 0.940000 & 0.860000 & 0.931000 \\
SCS-instr(gpt4omini) & 0.504000 & 0.531000 & 0.940000 & 0.860000 & 0.933000 \\
SCS-llm(gpt4omini) & 0.444000 & 0.470000 & 0.960000 & 0.880000 & 0.947000 \\
BM25 & 0.123000 & 0.143000 & 0.873000 & 0.773000 & 0.873000 \\
\label{tab:cord19_fulltext_trec_covid_train}
\end{longtable}

\begin{longtable}{lrrrrr}
\caption{Results for clinicaltrials\_2021\_trec\_ct\_2021} \\
\toprule
Ranker Name & KendallTau & Spearman & MRR & Prec@k=3 & MAP \\
\midrule
\endfirsthead
\caption[]{Results for clinicaltrials\_2021\_trec\_ct\_2021} \\
\toprule
Ranker Name & KendallTau & Spearman & MRR & Prec@k=3 & MAP \\
\midrule
\endhead
\midrule
\multicolumn{6}{r}{Continued on next page} \\
\midrule
\endfoot
\bottomrule
\endlastfoot
Bm25 PCS-instr(gpt4omini) & 0.190000 & 0.247000 & 0.958000 & 0.867000 & 0.830000 \\
Bm25 PCS-llm(gpt4omini) & 0.186000 & 0.240000 & 0.958000 & 0.876000 & 0.827000 \\
Bm25 SCS-instr(gpt4omini) & 0.190000 & 0.245000 & 0.980000 & 0.880000 & 0.829000 \\
Bm25 SCS-llm(gpt4omini) & 0.163000 & 0.210000 & 0.960000 & 0.867000 & 0.814000 \\
Bm25IdealTest & 0.235000 & 0.304000 & 0.951000 & 0.858000 & 0.849000 \\
Bm25IdealTrain & 0.239000 & 0.309000 & 0.951000 & 0.858000 & 0.851000 \\
bm25 & 0.205000 & 0.266000 & 0.949000 & 0.867000 & 0.837000 \\
\label{tab:clinicaltrials_2021_trec_ct_2021_test}
\end{longtable}

\begin{longtable}{lrrrrr}
\caption{Results for cord19\_fulltext\_trec\_covid} \\
\toprule
Ranker Name & KendallTau & Spearman & MRR & Prec@k=3 & MAP \\
\midrule
\endfirsthead
\caption[]{Results for cord19\_fulltext\_trec\_covid} \\
\toprule
Ranker Name & KendallTau & Spearman & MRR & Prec@k=3 & MAP \\
\midrule
\endhead
\midrule
\multicolumn{6}{r}{Continued on next page} \\
\midrule
\endfoot
\bottomrule
\endlastfoot
BM25 PCS-instr(gpt4omini) & 0.128000 & 0.162000 & 0.890000 & 0.787000 & 0.775000 \\
BM25 PCS-llm(gpt4omini) & 0.120000 & 0.153000 & 0.890000 & 0.787000 & 0.772000 \\
BM25 SCS-instr(gpt4omini) & 0.224000 & 0.279000 & 0.900000 & 0.807000 & 0.803000 \\
BM25 SCS-llm(gpt4omini) & 0.188000 & 0.235000 & 0.925000 & 0.820000 & 0.795000 \\
BM25IdealTest & 0.237000 & 0.296000 & 0.900000 & 0.793000 & 0.807000 \\
BM25IdealTrain & 0.233000 & 0.292000 & 0.910000 & 0.800000 & 0.807000 \\
BM25 & 0.199000 & 0.249000 & 0.912000 & 0.800000 & 0.796000 \\
\label{tab:cord19_fulltext_trec_covid_test}
\end{longtable}

\twocolumn

\section{Conclusion and Discussion}

In this paper, we introduce the problem and methods for \problem with a framework for generating custom datasets aimed at improving Query-By-Document (QBD) search. Our framework (Figure \ref{fig:qbd_dataset_gen}) leverages the capabilities of Large Language Models (LLMs) to incorporate domain-expert input, and produce document scores with explanations for human review. In this work, we specifically focus on the re-ranking step as it is the pivotal step in the computational process for \problem. The process we present can be used to generate datasets for QBD which can then be used to select or tune models for a particular use-case.

We presented a set of re-ranking methods, including single-document and pairwise-document scoring methods, and evaluated their performance with measures used to compare ranked lists including Kendall's $\tau_{b}$ Coefficient, Spearman’s rank-order correlation, and Mean Average Precision (MAP) on two TREC datasets. The results in these datasets showed that using LLM re-ranking alone to select or fine-tune models may not be sufficient, and can in fact hurt performance; adding the human-review step maybe necessary. This gives more reason to prefer using reranking methods that make human-review easier, such as those that do pairwise comparisons. 

In future work, we plan to extend the evaluation to include more datasets and additional LLM models in the reranking methods. We also plan to investigate the effect of varying the temperature in LLM models during the reranking step, and how to combine the document scores effectively after sampling scores with different temperatures. 

\section{Disclaimer}
This paper was prepared for informational purposes by
the Artificial Intelligence Research group of JPMorgan Chase \& Co. and its affiliates (``JP Morgan''),
and is not a product of the Research Department of JP Morgan.
JP Morgan makes no representation and warranty whatsoever and disclaims all liability,
for the completeness, accuracy or reliability of the information contained herein.
This document is not intended as investment research or investment advice, or a recommendation,
offer or solicitation for the purchase or sale of any security, financial instrument, financial product or service,
or to be used in any way for evaluating the merits of participating in any transaction,
and shall not constitute a solicitation under any jurisdiction or to any person,
if such solicitation under such jurisdiction or to such person would be unlawful.

\bibliography{main}

\end{document}